# Strain-relaxed GeSn-on-Insulator (GeSnOI) microdisks

**Daniel Burt,**[1,†] **Hyo-Jun Joo,**[1,†] **Yongduck Jung,**[1] **Youngmin Kim,**[1] **Yi-Chiau Huang,**[2] **and Donguk Nam**[1,*]

[1] *School of Electrical and Electronic Engineering, Nanyang Technological University, 50 Nanyang Avenue, Singapore 639798, Singapore*
[2] *Applied Materials Inc., Sunnyvale, CA 95054-3299, USA*
[†]*These authors contributed equally to this work.*
*\*dnam@ntu.edu.sg*

**Abstract:** GeSn alloys offer a promising route towards a CMOS compatible light source and the realization of electronic-photonic integrated circuits. One tactic to improve the lasing performance of GeSn lasers is to use a high Sn content, which improves the directness. Another popular approach is to use a low to moderate Sn content with either compressive strain relaxation or tensile strain engineering, but these strain engineering techniques generally require optical cavities to be suspended in air, which leads to poor thermal management. In this work, we develop a novel dual insulator GeSn-on-insulator (GeSnOI) material platform that is used to produce strain-relaxed GeSn microdisks stuck on a substrate. By undercutting only one insulating layer (i.e., Al$_2$O$_3$), we fabricate microdisks sitting on SiO$_2$, which attain three key properties for a high-performance GeSn laser: removal of harmful compressive strain, decent thermal management, and excellent optical confinement. We believe that an increase in the Sn content of GeSn layers on our platform can allow achieving improved lasing performance.

## 1. Introduction

Silicon (Si) photonics offers an elegant solution to overcome the current interconnect bottleneck facing modern integrated circuits (ICs), by replacing metallic interconnections with optical counterparts [1]. However, the lack of a complementary metal-oxide-semiconductor (CMOS) compatible light source is the limiting factor for achieving such optical interconnects. Si is an indirect bandgap material and therefore not suitable for the generation of light [2]. Despite being an indirect bandgap material like Si, germanium (Ge) is a promising candidate for a CMOS compatible light source [2,3]. The energy difference between the direct Γ and indirect L conduction valleys in Ge is relatively small (~140 meV), therefore Ge can be converted into a direct bandgap using various techniques. Tensile strain engineering can be used to reduce the energy difference between the direct Γ and indirect L conduction valleys, which improves light emission [3,4]. Both optically and electrically pumped lasing in Ge were achieved using a small amount of tensile strain and heavy n-type doping [3,5,6]. Despite this remarkable achievement, the injection threshold power and current densities were far too high for practical applications. Adding large mechanical tensile strain can further reduce the energy difference between the direct Γ and indirect L conduction valleys [7–12], which led to the observation of optically pumped lasing operating at a cryogenic temperature [13–15].

Alloying Ge with tin (Sn) to make GeSn is another promising approach to achieve a CMOS compatible light source [16]. Sn alloying can convert GeSn into a direct bandgap semiconductor at an Sn content of around 6-8 at% [17,18]. Proof-of-concept lasing was first demonstrated in GeSn alloys with an Sn content of 12.6 at% [19], which motivated a number of research groups to improve the performance of GeSn lasers. The research has predominantly focused on increasing the Sn content to improve the directness for reducing the threshold power density and increasing the operating temperature, which resulted in many successful demonstrations [20–25]. By increasing the Sn content to 20 at%, for example, the operating temperature was increased to 270 K that is close to room temperature [25]. Unfortunately, the threshold was still quite high (886 kW cm$^{-2}$), which requires further investigation on how to achieve both low threshold and high operating temperature. The main issue with increasing Sn content is a higher dislocation density partly due to an increased lattice mismatch with the underlying Ge buffer layer and also because of the lower growth temperatures required due to a decreased thermal stability [26]. Another major issue with increasing the Sn content is that a larger compressive strain is introduced into the GeSn layer, which works against the Sn alloying by reducing the directness [17,18,25,27,28].

To solve the issue of the harmful compressive strain in higher Sn content layers, researchers utilized partially suspended structures [22,23,29,30]. By releasing the underlying Ge layer, the GeSn gain medium can be fully relaxed, achieving an improved directness in the suspended regions where optical modes are located. Recent demonstrations have shown that lasing properties can be improved when low Sn content alloying is combined with tensile strain engineering [31–33], presenting an ultra-low threshold of 0.8 kW cm$^{-2}$ for an Sn content of 5.5 at% and a biaxial tensile strain of 1.4%. The combination of tensile strain and low Sn content alloying also allowed for the first demonstration of the continuous-wave operation of an optically pumped laser in GeSn alloys [32]. All the previously described lasing demonstrations with either relaxed or tensile strained GeSn layers require the structures to be suspended for achieving both high optical confinement and strain-engineered gain media. Unfortunately, suspended structures are more susceptible to heating, which increases the optical losses and inhibits lasing at higher temperatures. It is highly desirable to achieve a platform with good optical confinement, decent thermal conduction (i.e., no suspension), and strain relaxation.

In this article, we demonstrate a fully CMOS-compatible GeSnOI platform that can achieve all three criteria. Using a low-temperature direct bonding technique, we obtain a high-quality GeSn layer on dual insulators by removing the highly defective Ge/GeSn interface, as evidenced by transmission electron microscopy (TEM). Microdisk optical cavities were fabricated on this platform and subsequently relaxed by undercutting the Al$_2$O$_3$ layer, thus releasing the harmful compressive strain in the GeSn gain medium, as confirmed by 2D Raman mapping. The novel dual insulator GeSnOI platform enables the first demonstration of strain-relaxed microdisk cavities stuck to SiO$_2$. Pump-power-dependent Raman studies were conducted at room temperature to demonstrate the superior thermal management of the stuck microdisks compared to the suspended counterparts. Comprehensive photoluminescence (PL) studies were conducted on the microdisks to provide further evidence for the excellence of our platform. Temperature-dependent PL studies show improved PL emission from the stuck microdisks compared to the bulk case, which is attributed to the increased directness in strain-relaxed GeSn layers. Lastly, pump-power-dependent PL studies conducted at low temperatures further confirmed the superior thermal management of the stuck microdisks. Our work presents a new GeSn platform that can be used for creating more efficient GeSn lasers.

## 2. Bonding and device fabrication procedures

### 2.1) Bonding procedure

A 200 nm Ge$_{0.94}$Sn$_{0.06}$ epilayer was grown on a 12-inch Si wafer with a 1000-nm Ge buffer layer using low-pressure chemical vapor deposition (LPCVD). This 12-inch wafer was then diced into 3cm-by-3cm dies to provide the carrier chips. Thermal oxide with a thickness of 1000 nm was grown on a 6-inch Si wafer and then diced into 3cm-by-3cm dies to provide the handle chips. Both handle and carrier chips were cleaned with ammonium hydroxide solution (10 wt%) and loaded into an atomic layer deposition (ALD) chamber, after which a 150-nm Al$_2$O$_3$ layer was deposited at 250°C using trimethylaluminum (TMA) and water as precursors. The Al$_2$O$_3$ layer acts as both a sacrificial layer and as a bonding layer [34]. Outgas annealing was then performed at 350°C in a tube furnace under a nitrogen gas ambient to drive out any residual gasses trapped in $Al_2O_3$ during the deposition process (mostly hydrogen gas), which could cause debonding in the post bonding annealing step. To ensure a smooth surface for both the carrier and handle chips that is a strict requirement for direct bonding, both chips underwent a chemical mechanical planarization (CMP) step, which reduced the surface roughness to 0.2 nm as confirmed by atomic force microscopy (AFM). Post CMP cleaning using ammonium hydroxide solution (10 wt%) and a polyvinyl alcohol (PVA) brush, removed any debris and slurry particles from the CMP process, leaving a clean surface for bonding. Both handle and carrier chips were rinsed with de-ionized (DI) water to promote good surface hydroxyl coverage and dried using a nitrogen gun. Both the handle and carrier chip surfaces were brought into direct contact to initiate the bond wave, and Van der Waals forces were formed between the two chips. Post bonding annealing was conducted in a tube furnace under a nitrogen gas ambient at 350°C for 3 hours to increase the bonding strength by forming covalent bonds. Mechanical grinding was then performed on the bonded chips to thin down the top carrier Si to a thickness of 50 μm, followed by wet etching using 30 wt% potassium hydroxide (KOH) at 80°C, which selectively removed the remaining Si while leaving the Ge and $Ge_{0.94}Sn_{0.06}$ layers intact.

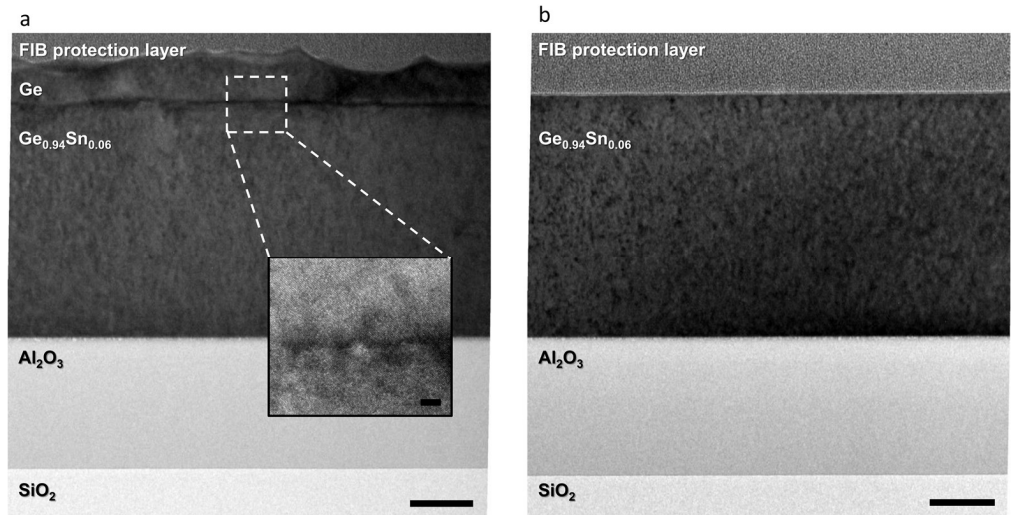

Fig. 1. (a) Cross-sectional TEM image with a thin layer of the remaining Ge buffer layer, showing the highly defective Ge/$Ge_{0.94}Sn_{0.06}$ interface. Scale bar, 50 nm. Inset scale bar, 5 nm. (b) Cross-sectional TEM image after the removal of the defective interface. Scale bar, 50 nm.

A final CMP step removed the Ge buffer layer and the highly defective interface, leaving a smooth and high-quality $Ge_{0.94}Sn_{0.06}$ layer with a thickness of ~200 nm. To illustrate the removal of the highly defective interface between Ge and $Ge_{0.94}Sn_{0.06}$, we show TEM cross-sectional images before (Fig. 1(a)) and after (Fig. 1(b)) removing the highly defective Ge and $Ge_{0.94}Sn_{0.06}$ interface. The final material stack consists of $Ge_{0.94}Sn_{0.06}$, $Al_2O_3$, $SiO_2$, and Si (not shown). The $Al_2O_3$ layer is used as a sacrificial layer to release the harmful compressive strain in GeSn while the $SiO_2$ layer allows achieving excellent optical confinement and decent thermal management simultaneously.

*2.2) Device fabrication procedure*

The dual insulator GeSnOI samples were diced into 1cm-by-1cm chips for device fabrication, followed by solvent cleaning using acetone and isopropanol alcohol (IPA). Positive photoresist AZ5214E was spin-coated onto the chip at 6000 RPM. Photolithography was used to define the microdisks, followed by pattern transfer using $Cl_2$- and $BCl_3$-based inductively coupled plasma (ICP) dry etching to etch down to the $Al_2O_3$ sacrificial layer and form the microdisks. The chip was then cleaved into three, and one chip underwent no further processing to create the 'bulk' microdisks. The other two chips underwent wet etching using 30 wt% KOH at 80°C, which was used to laterally undercut 4 μm of the $Al_2O_3$ layer to produce the stuck and suspended microdisks. Microdisks were stuck to the underlying $SiO_2$ utilizing the capillary forces introduced during the drying step after rinsing the sample in DI water. The suspended structures were made by rinsing in IPA with a lower surface tension, thereby reducing capillary forces during the drying process and preventing the microdisk being stuck to the underlying $SiO_2$. Figure 2(a) shows a schematic process flow of the device fabrication process. Figures 2(b) and (c) show scanning electron microscopy (SEM) images of suspended and stuck microdisks, respectively. The microdisks investigated in this study all had a diameter of 9 μm to allow for a fair comparison.

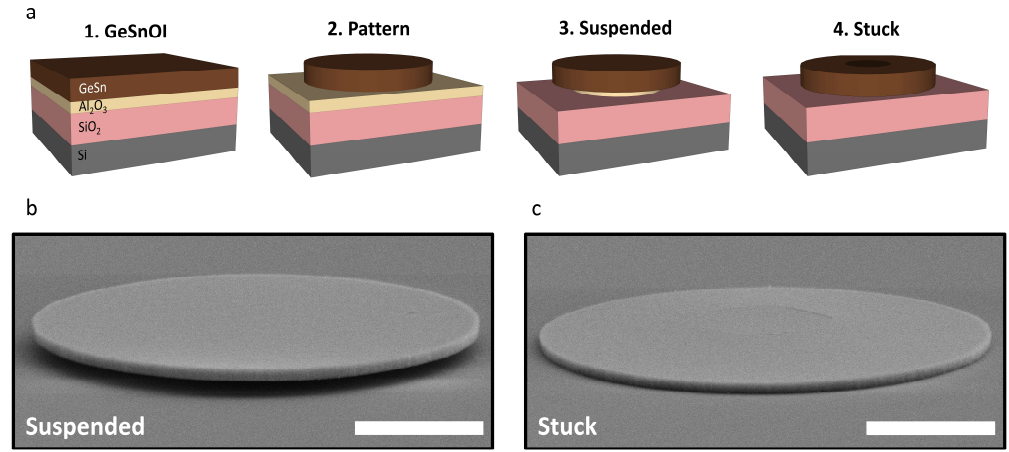

Fig. 2. (a) Schematic of the fabrication process. Tilted SEM images of (a) suspended and (b) stuck microdisks. Scale bar, 4 μm.

## 3. Raman measurement and analysis

To confirm the relaxation of the compressive strain in the released microdisks, Raman spectroscopy and 2D Raman mapping were carried out. A 532-nm laser was focused onto the microdisks by a ×100 objective lens, which produced a spot size of ~1 μm. Pump-power-dependent measurements were conducted to select a suitable power for the mapping and ensure that no heating effects occurred [35] as heating can cause the relative Raman peak position to shift to lower values and therefore cause an error in measuring the strain relaxation.

Figures 3(a) and (b) show the 2D Raman maps for the bulk (i.e., not released) and stuck microdisks, respectively. The bulk microdisk shows a fairly uniform Raman peak value of ~297.5 cm$^{-1}$. The stuck microdisk had a similar value to the bulk microdisk in the central region, as this region was not undercut, and the strain was not relaxed. However, for the outer region where $Al_2O_3$ was undercut, the Raman peak value has shifted to ~295.5 cm$^{-1}$, thus confirming the relaxation of the compressive strain. A slightly larger Raman peak shift to ~294.5 cm$^{-1}$ is observed at the interface between the central non-undercut region and outer undercut region for the stuck microdisk. This additional tensile strain can be attributed to the bending of the GeSn layer occurring when the outer region of the microdisk is stuck to $SiO_2$.

Figure 3(c) shows the Raman spectra of the bulk and the stuck microdisks at the circumference. The Raman spectrum of bulk single-crystal Ge is also shown as a reference. The spectra were fitted using the Lorentzian function to accurately determine the Raman peak position. The peak position of the bulk microdisk was located at 297.55 cm$^{-1}$, whereas the peak position of the stuck microdisk is lower at 295.35 cm$^{-1}$ due to the relaxation of the intrinsic compressive strain. The Raman shift from the compressively strained to relaxed state was 2.2 cm$^{-1}$, which corresponds to an intrinsic compressive strain of ~0.4%, calculated using a Raman-strain coefficient of 521 cm$^{-1}$ [36]. The relaxation of compressive strain results in an approximate increase the directness of GeSn by 50 meV in the stuck microdisk according to theoretical calculations [18].

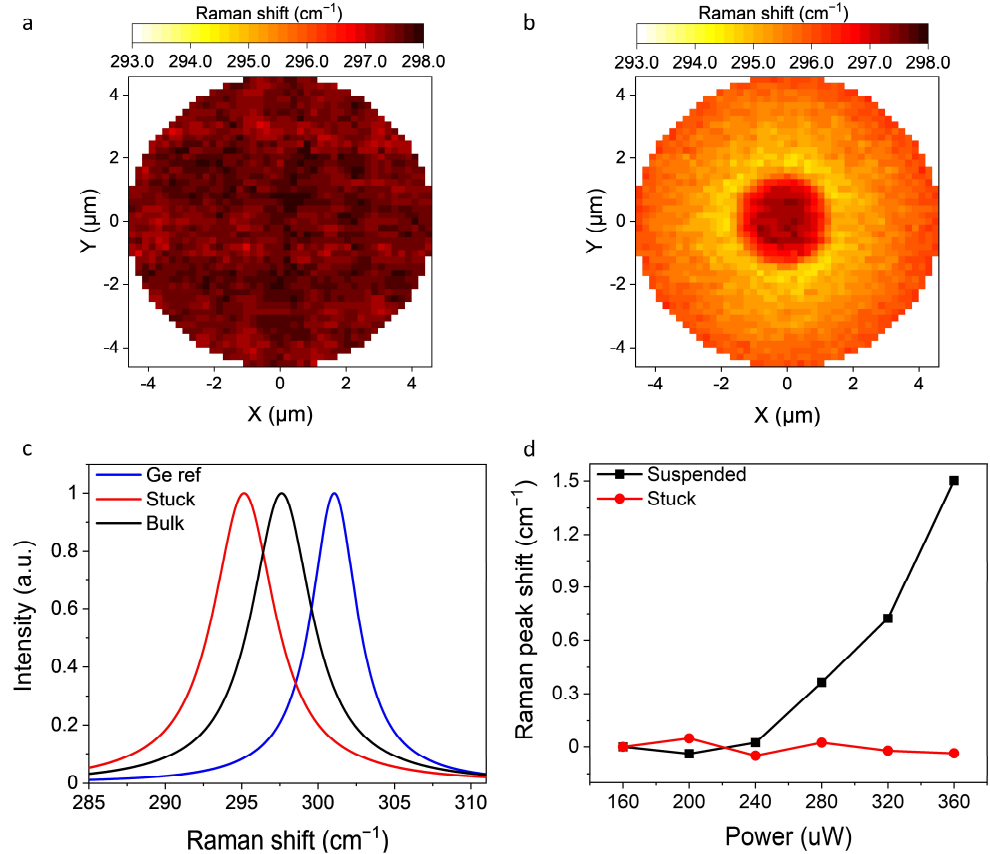

Fig. 3. 2D Raman maps of the (a) bulk and (b) stuck microdisks. (c) Raman spectra of the bulk and released microdisks with Lorentzian fitting (Ge for reference). (d) Pump-power-dependent Raman peak shifts for the suspended and stuck microdisks.

To confirm whether the stuck microdisk has improved thermal management compared to the suspended microdisk, pump-power-dependent Raman spectroscopy was conducted. Figure 3(d) shows the shift of the relative Raman peak position as a function of the pump power for both the stuck and suspended microdisk, with the beam focused on the circumference of the microdisks. For the stuck microdisk, the relative Raman peak position does not shift as the pump power is increased. However, for the suspended microdisk, there is a sizeable heat-induced shift of the Raman peak position (~1.5 cm$^{-1}$) at the maximum pump power of 360 μW. The obvious difference in the pump-induced Raman shift illustrates the superior thermal management of the stuck microdisk compared to the suspended counterpart owing to the higher thermal conductivity of SiO$_2$ compared to air.

## 4. PL measurement and analysis

We performed comprehensive PL measurements and analyses to investigate the effect of strain relaxation and thermal management on the optical emission properties in the microdisks. We utilized a 1,550-nm pulsed laser with a pulse width and repetition rate of 50 ns and 3 MHz, respectively. The samples were placed into a helium cryostat operating between 4 K and 300 K. The pulsed 1,550-nm pump laser was focused onto the sample using a ×15 reflective objective lens producing a spot size of roughly 10 μm. The signal from the sample was collected by the same reflective objective lens and subsequently coupled into a grating which diffracted the spectrum onto a 1D-array extended InGaAs detector with a wavelength detection range between 1.4 and 2.1 μm.

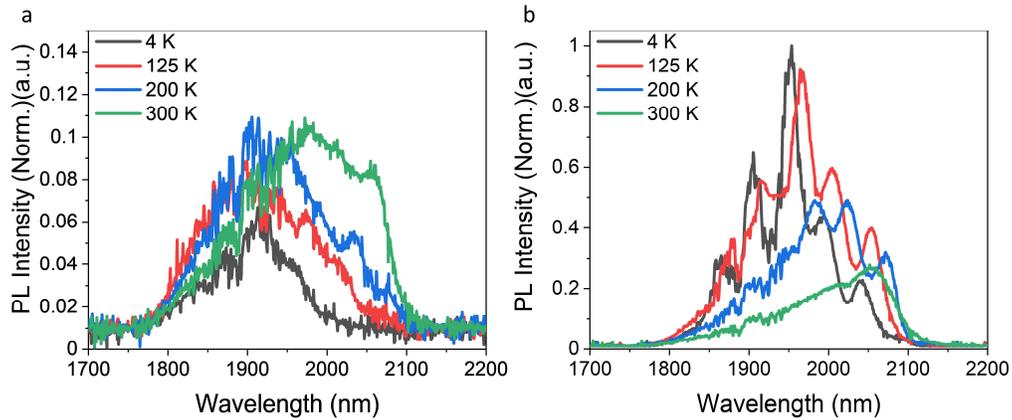

Fig. 4. Temperature-dependent PL from the (a) bulk and (b) stuck microdisks.

Temperature-dependent measurements were taken at a fixed pump power of 25 mW for both the bulk and the stuck microdisks to study the effect of relaxing the compressive strain on the optical emission. Figures 4(a) and (b) show the PL spectra at various temperatures for the bulk and the stuck microdisks, respectively. All the spectra in Figs. 4(a) and (b) are normalized to the spectrum of the stuck microdisk taken at 4 K (dark grey). It should be noted that the y-axis scales for Figs. 4(a) and (b) are different. For the bulk microdisk at 4 K, only broad spontaneous emission was observed. For the stuck microdisk, in contrast, clear resonances from the whispering gallery modes (WGMs) can be observed. These WGMs can be seen up to 200 K but disappear at 300 K due to increased material losses at higher temperatures. The spectra for the bulk microdisk had no obvious WGMs at any temperature, and the intensity was around an order of magnitude lower than the stuck microdisk. Both the bulk and the stuck microdisks were in contact with the underlying oxides and should have comparable thermal management, and the material quality is constant; therefore, the larger intensity and the presence of clear WGMs in the stuck microdisks should arise from the improved directness, which originates from the relaxation of the compressive strain in the stuck microdisk. It should be noted that the optical resonances were relatively weak, which may be partly due to the thin GeSn layer (~200 nm), for which optical confinement is not ideal, and also due to the high material loss arising from the low Sn content in the active medium. The aim of this study is to provide a proof-of-concept demonstration presenting that strain relaxation could be achieved in the dual insulator GeSnOI platform, which can also achieve high optical confinement and decent thermal management. Future studies will focus on increasing the Sn content with thicker layers to achieve optical gain.

Pump-power-dependent PL measurements were conducted at 4 K to compare the effect of the high pump power between the stuck and suspended microdisks. Figures 5(a) and (b) show the PL spectra with increasing pump power for the suspended and stuck microdisks, respectively. All the spectra in Figs. 5(a) and (b) are normalized to the spectrum of the stuck microdisk taken with 40 mW (green). The y-axis scales for Figs. 5(a) and (b) are different. As the pump power increases for the suspended microdisk, the emission redshifts due to heating effects [33]. Furthermore, the WGM resonance peaks become broader as significant heating effects increase material losses in GeSn. For the stuck microdisk, in contrast, the resonance peaks slightly blueshift with increasing pump power, and there is no visible degradation of the optical modes, indicating no observable heating effects. The lack of broadening of the optical resonances and no observable red shifting of the emission confirms the superior thermal management of the stuck microdisk compared to the suspended case, which was enabled by our dual insulator GeSnOI. Finite-element method (FEM) simulations confirmed that under the optical pumping conditions used in our PL experiments, the temperature of the suspended microdisk increases by ~60 K, whereas the stuck microdisk shows a negligible temperature increase.

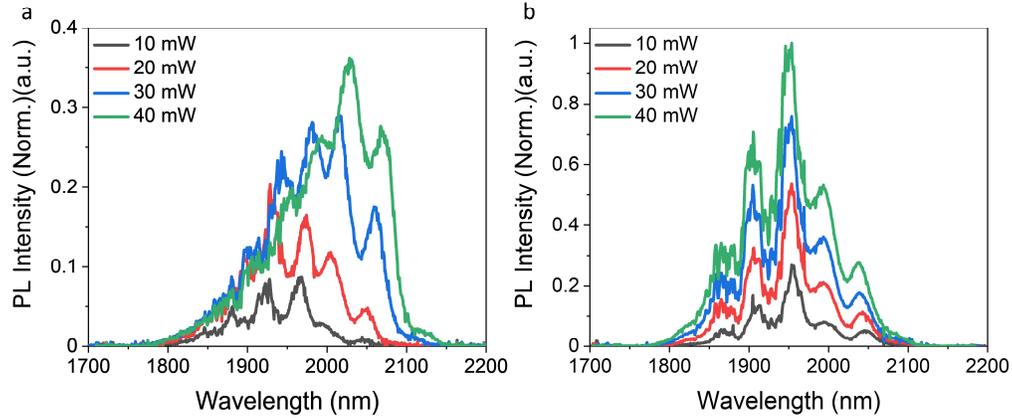

Fig. 5. Pump-power-dependent PL from the (a) suspended and (b) stuck microdisks at 4 K.

## 5. Conclusion

In this paper, we have presented a novel dual insulator GeSnOI platform that achieves strain relaxation, good optical confinement, and decent thermal management simultaneously. Furthermore, we demonstrated the removal of the highly defective Ge/GeSn interface as evidenced by TEM analyses. Raman spectroscopy was used to confirm the strain relaxation in the microdisk stuck to $SiO_2$, showing superior thermal management compared to the suspended microdisk. Comprehensive temperature-dependent PL studies on the bulk and stuck microdisks were performed to investigate and compare the optical emission properties between the stuck and bulk microdisks, showing improved optical emission for the stuck microdisk originating from the increased directness. Pump-power-dependent low-temperature PL studies on the stuck and suspended microdisks further confirmed that the stuck microdisk achieves superior thermal management compared to the suspended microdisk. Our new dual insulator GeSnOI platform may provide unforeseen opportunities to push the performance of GeSn lasers to the limit when GeSn layers with higher Sn contents are employed.

**Funding**

Ministry of Education, Singapore, under grant AcRF TIER 1 2019-T1-002-050 (RG 148/19 (S)); Ministry of Education, Singapore, under grant AcRF TIER 2 (MOE2018-T2-2-011 (S)); National Research Foundation of Singapore through the Competitive Research Program (NRF-CRP19-2017-01); National Research Foundation of Singapore through the NRF-ANR Joint Grant (NRF2018-NRF-ANR009 TIGER); the iGrant of Singapore A*STAR AME IRG (A2083c0053).

**Disclosures**

The authors declare no conflicts of interest.

**Data availability**

Data underlying the results presented in this paper are not publicly available at this time but may be obtained from the authors upon reasonable request.